\documentclass[acmsmall]{acmart}
\AtBeginDocument{%
  \providecommand\BibTeX{{%
    \normalfont B\kern-0.5em{\scshape i\kern-0.25em b}\kern-0.8em\TeX}}}

\usepackage{multirow}
\usepackage{booktabs}
\usepackage{tcolorbox}
\usepackage{adjustbox}
\usepackage{amsmath,amsfonts}
\usepackage{booktabs}
\usepackage[font={footnotesize}]{caption}
\usepackage{color, colortbl}
\usepackage{dblfloatfix}
\usepackage{enumitem}
\usepackage{graphicx}
\usepackage{cleveref} 
\usepackage{listings}
\usepackage{mdframed}
\usepackage{multirow}
 
\usepackage{pdfcomment}
\usepackage{relsize}

\usepackage{soul}
\usepackage{subcaption}
\usepackage{tabularx}
\usepackage{textcomp}
\usepackage{wasysym}
\usepackage{xcolor}
\usepackage{xspace}
\usepackage{float}

\def\Snospace~{\S{}}

\usepackage{tikz}
\usepackage{pgfplots}
\pgfplotsset{compat=1.13}
\usetikzlibrary{calc}
\usetikzlibrary{positioning,arrows,fit,matrix}
\usetikzlibrary{backgrounds}
\usetikzlibrary{patterns}
\usetikzlibrary{shapes.arrows} %
\usepgfplotslibrary{groupplots}
\usepgfplotslibrary{statistics}
\usepgfplotslibrary{fillbetween}

\tikzset{
	font=\footnotesize
}

\definecolor{cbone}  {HTML}{006BA4} %
\definecolor{cbtwo}  {HTML}{FF800E} %
\definecolor{cbthree}{HTML}{ABABAB} %
\definecolor{cbfour} {HTML}{595959} %
\definecolor{cbfive} {HTML}{5F9ED1} %
\definecolor{cbsix}  {HTML}{C85200} %
\definecolor{cbseven}{HTML}{898989} %
\definecolor{cbeight}{HTML}{A2C8EC} %
\definecolor{cbnine} {HTML}{FFBC79} %
\definecolor{cbten}  {HTML}{CFCFCF} %

\definecolor{applegreen}{rgb}{0.55, 0.71, 0.0}
\definecolor{goeblue}{RGB}{0,51,102}
\definecolor{tubsredSec}{cmyk}{0.0,1.00,0.6,0.6}
\definecolor{tubsredPrim}{cmyk}{0.1,1.0,0.8,0.0}

\mdfdefinestyle{MyFrame}{
	linecolor=black,
	backgroundcolor=gray!10,
	innertopmargin=5pt,
	innerbottommargin=10pt,
	innerrightmargin=5pt,
	innerleftmargin=5pt,
}
\newcommand{\pitfallenv}[3]{%
	\vspace{2mm}
	\noindent\begin{minipage}{\columnwidth}%
	\begin{mdframed}[style=MyFrame]
		\textbf{#1 -- #2.} #3
	\end{mdframed}%
\end{minipage}%
}
\newcommand{\hazard}[5]{%
	\vspace{2mm}
	\noindent\begin{minipage}{\columnwidth}%
	\begin{mdframed}[style=MyFrame]
		\textbf{#1 -- #2} (\emph{#3}) \\
            #4
	\end{mdframed}%
\end{minipage}%
}

\setcopyright{acmcopyright}
\copyrightyear{2018}
\acmYear{2018}
\acmDOI{XXXXXXX.XXXXXXX}

\acmJournal{JACM}
\acmVolume{37}
\acmNumber{4}
\acmArticle{111}
\acmMonth{8}

\begin{document}

\title{Hazards in Deep Learning Testing: Prevalence, Impact and Recommendations}


\author{Salah Ghamizi}
\affiliation{%
  \institution{University of Luxembourg}
  \streetaddress{29 Av. John F. Kennedy}
  \city{Kirchberg Luxembourg}
  \country{Luxembourg}}
\email{salah.ghamizi@uni.lu}

\author{Maxime Cordy}
\affiliation{%
  \institution{University of Luxembourg}
  \streetaddress{29 Av. John F. Kennedy}
  \city{Kirchberg Luxembourg}
  \country{Luxembourg}}
\email{maxime.cordy@uni.lu}

\author{Yuejun Guo}
\affiliation{%
  \institution{Luxembourg Institute of Science And Technology}
  \streetaddress{5 Av. des Hauts-Fourneaux}
  \city{Esch-Belval Esch-sur-Alzette}
  \country{Luxembourg}}
\email{yuejun.guo@list.lu}

\author{Mike Papadakis}
\affiliation{%
  \institution{University of Luxembourg}
  \streetaddress{29 Av. John F. Kennedy}
  \city{Kirchberg Luxembourg}
  \country{Luxembourg}}
\email{michail.papadakis@uni.lu}

\author{Yves Le Traon}
\affiliation{%
  \institution{University of Luxembourg}
  \streetaddress{29 Av. John F. Kennedy}
  \city{Kirchberg Luxembourg}
  \country{Luxembourg}}
\email{yves.letraon@uni.lu}
\renewcommand{\shortauthors}{Trovato and Tobin, et al.}

\begin{abstract}
Much research on Machine Learning testing relies on empirical studies that evaluate and show their potential. However, in this context empirical results are sensitive to a number of parameters that can adversely impact the results of the experiments and potentially lead to wrong conclusions (Type I errors, i.e., incorrectly rejecting the Null Hypothesis). To this end, we survey the related literature and identify 10 commonly adopted empirical evaluation hazards that may significantly impact experimental results. We then perform a sensitivity analysis on 30 influential studies that were published in top-tier SE venues, against our hazard set and demonstrate their criticality. Our findings indicate that all 10 hazards we identify have the potential to invalidate experimental findings, such as those made by the related literature, and should be handled properly. Going a step further, we propose a point set of 10 good empirical practices that has the potential to mitigate the impact of the hazards. We believe our work forms the first step towards raising awareness of the common pitfalls and good practices within the software engineering community and hopefully contribute towards setting particular expectations for empirical research in the field of deep learning testing. 
\end{abstract}



\keywords{Do, Not, Us, This, Code, Put, the, Correct, Terms, for,
  Your, Paper}

\received{20 February 2007}
\received[revised]{12 March 2009}
\received[accepted]{5 June 2009}

\maketitle

\section{Introduction}
\label{sec:intro}


Machine learning (ML) research and practice has established empirical criteria and methods for the adequate deployment of ML systems. 
Recent research raised awareness about the importance of robustness \cite{Carlini2019OnEA}, privacy \cite{liu2020privacy}, fairness \cite{du2020fairness}, interpretability \cite{li2022interpretable} and efficiency \cite{menghani2023efficient}.  

Inspired by the software testing research, the last few years have showed a growing interest in machine learning testing. The latter however includes challenges that spawn from the inherent properties of machine learning systems and are in contrast to the traditional software systems~\cite{zhang2020machine}. Indeed, concepts like test oracles, regressions, and coverage are hardly applicable to ML testing. Thus, the use of software testing methods may lead to overly optimistic evaluations and, worse, may impact ML workflows and undermine basic assumptions, research conclusions, and recommendations.  

The immaturity of the field makes it hard for researchers to follow a sound scientific methodology that can fundamentally support the key intuitions and experimental conclusions. The key issues here are two; the lack of experience and the sensitivity of the empirical studies to a number of parameters and assumptions that are incompatible with traditional software testing. We thus, argue that a set of pitfalls, principles and good practices need to be assessed and addressed in any work tackling ML testing, and even more when involving deep learning (DL) models. 

In this paper, we identify 10 common, yet subtle, evaluation hazards that may compromise the validity of an empirical study on DL testing and make it difficult to draw reliable conclusions and comparisons with previous related work. Across the 10 hazards, we propose 10 good practices to mitigate their impact with increasing cost and complexity.

To support this claim, we analyze the prevalence of these hazards in 30 Top Software Engineering papers from the past five years that propose techniques and frameworks for DL testing. As we might suspect, each paper insufficiently addressed at least three of the hazards. Even worse, some hazards have not been addressed by any of the papers. 

To understand the impact of these hazards, and how they can compromise the results of the studies that are not addressing them, we perform an impact analysis study of each of the hazards and evaluate the severity of each of them.  

In summary, the paper makes the following contributions:
\begin{enumerate}
    \item \textbf{Scope Definition.} We revisit the DL testing workflow to highlight where the evaluation hazards may arise.  
    \item \textbf{Hazards Identification.} We identify ten hazards in the common practices of DL testing. We also propose a set of actionable  strategies, with progressive costs, to correctly assess their influence and mitigate their impacts on the conclusions of DL testing papers.
    \item \textbf{Prevalence Analysis.} Starting from 86 research papers, we focused the scope of our study on 30 SE papers and evaluated the prevalence of each hazard.
    \item \textbf{Impact Analysis.} For each critical hazard, we conducted evaluation experiments on the six most popular DL testing libraries, and demonstrate how a different the hazards change the paper conclusions. 
\end{enumerate}
\section{ML Testing}
\label{sec:scope}


The 2020's survey of \citet{zhang2020machine} is the first systematic attempt to lay a common ground for ML testing concepts, workflow and literature. This work has played an important role in structuring (and arguably driving) research work in ML testing. Along the lines of this seminal work, we aim to provide an updated perspective of ML testing by identifying experimental pitfalls that may invalidate (or tone down) experimental conclusions and hinder the progress and adoption of such research advances.

\subsection{An updated definition}

Our starting point is the very definition of ``ML testing'' given in this paper (and later more broadly adopted by the SE literature), i.e. \emph{``any activity designed to reveal machine learning bugs''}, where an ML bug is \emph{``any imperfection in a ML item that causes a discordance between the existing and required conditions''}. Applied in particular to classification tasks, this definition entails that ML testing are activities discovering discordance between the classification prediction of an ML model and this ground truth label of the input.

The goal of revealing bugs is an inheritance of the (classical) software testing view where all discovered bugs should (ideally) be fixed, making software testing the initial step to improving software reliability. However, a blunt transposition of this definition to ML models would disregard the specifics of these systems compared to classical software and, in turn, run the risk of hindering the capability of testing methods to improve ML model quality. 

Indeed, the large amount of data and the associated uncertainty of ML models makes correctness a non-absolute requirement. This means that ML bugs are tolerable as long as their proportion remains within a given statistical envelope. More generally, ML models typically deal with high-dimension data space, where for any input there exist a large number of $\epsilon$-close neighbouring inputs. For ML testing, this implies that near an ML bug lie a large number of similar ML bugs exerting the same defect. The utility of revealing bugs thus decreases as discovered bugs are similar. Nevertheless, bug fixing is not achieved by an exogenous process code modifying source code but rather by an endogenous process that retrains the model using additional data (often, the generated tests). Thus, any revealed ML bugs can (and should) directly be used to improve the model. Thus, overall, ML testing has to deal with (1) many bugs which (2) can be used to improve the model but (3) not all are useful. These characteristics lead us to dismiss bug discovery as the sole and central objective of ML testing, and rather make a new definition that emphasizes the importance of improving the model.

\begin{definition}[ML Testing]
Machine Learning Testing (ML testing) refers to any activity designed to \textbf{reveal or engineer} machine learning bugs that could be \textbf{leveraged to improve} the machine learning model under evaluation.
\end{definition}

A striking illustration of an improvement-driven ML testing approach is the \emph{adversarial example} phenomenon. Research on ML robustness came up with simple methods to create ML bugs through the introduction of carefully chosen perturbations to benign inputs, leading to incorrect predictions. Automated methods (``adversarial attacks'') can create a large number of such examples with high success rate. Then, ``Adversarial hardening'' is the process of using adversarial examples to improve model robustness against attacks. 

\subsection{Workflow}

\begin{figure}
    \centering
    \vspace{-0.6em}
    \includegraphics[width=\linewidth]{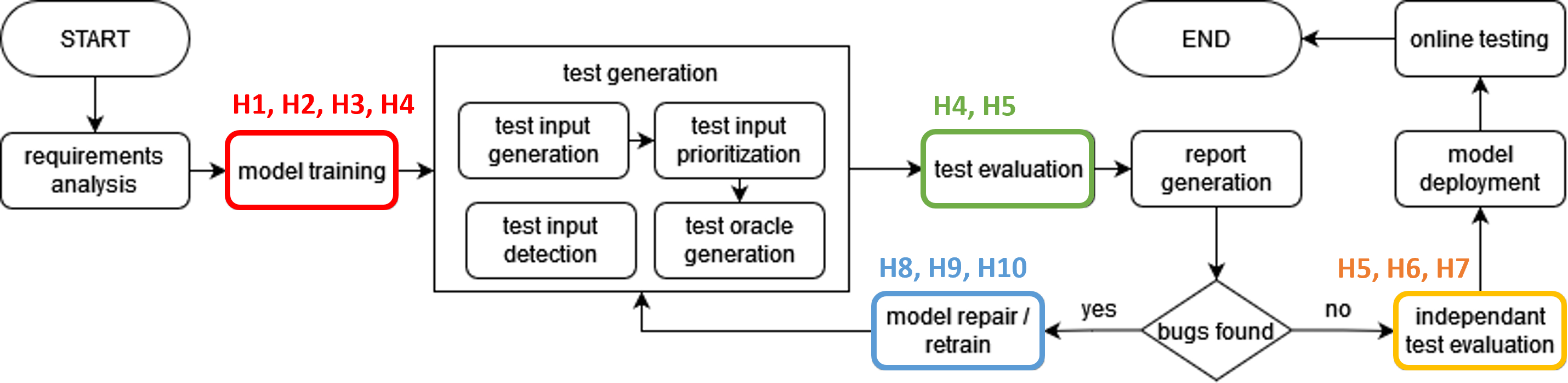}
    \vspace{-2.25em}
    \caption{Proposed ML pipeline. Its an extends \citet{zhang2020machine} pipeline by considering the input selection and detection, and by adding an independent test evaluation. We highlight the steps where each hazard occurs}
    \vspace{-0.75em}
    \label{fig:ml_pipeline}
\end{figure}

To scope our study, we refine the ML workflow proposed by \citet{zhang2020machine} as shown in Figure \ref{fig:ml_pipeline}. ML testing is commonly split between \emph{offline}, where the model is assessed with historical data before its deployment, and  \emph{online} where the model is confronted to real-time or quasi-real-time test inputs. The focus of our work is offline testing, thus we do not detail the online testing part and refer to the original workflow for more details. It is to be noted, however, that some techniques meant for offline testing can be used and evaluated in online settings.

Our pipeline shares some components with the pipeline proposed by Zhang et al.\cite{zhang2020machine}. In this idealised scenario, an initial model is trained following the requirements of the problem. The \textbf{requirements' analysis} will guide the selection of the initial tasks of the model and its design choices, eg, the architectures, its hyper-parameters, the loss, optimizer, and training augmentations. These design choices need to remain the same between the initial \textbf{model training} and the recursive \textbf{model repair/retrain} steps of the pipeline.
Next, a relevant test set is generated and evaluated. The test generation involves several steps, and our literature review identified four main steps. In the following, we will mostly focus on the literature of image classification machine learning.

\textbf{Test input generation}: This step builds a set of inputs that exhibit interesting properties for the evaluation of the model. These inputs can be collected in the wild from the same or a close distribution to the original training set. They can also be totally artificial, using for example generative models (GANs, AutoEncoders, etc). The test inputs can also be hybrid (perturbed real inputs \cite{pei2017deepxplore,goodfellow2014explaining,zhang2017mixup,yun2019cutmix}). 

\textbf{Test oracle generation}: Depending on the "similarity" of the hybrid inputs to their original ones, the associated oracles, i.e., labels for classification tasks, can be inferred from the original oracles. In case of perturbations, e.g., adversarials, the oracles are generally assumed to be the same to the original ones. In case of input mixing, the oracles are weighted combinations of the original ones.

\textbf{Test input prioritization}: The generated inputs can be too large to infer their oracles, or to evaluate. 
Then, an optional step is to prioritize the inputs for which the oracles will be generated. The selection metrics can be model dependent\cite{feng2020deepgini,kim2018guiding}, or independent\cite{shao2019learning}.

\textbf{Test input detection:} Given inputs with associated labels, an optional step is to detect/infer the relevance of the input without the evaluation of their performance on the target model. 
These detectors can be plugged in the input generation process to drive the generation towards inputs that optimize the detected behaviour. 

\textbf{Test evaluation} and \textbf{independent test evaluation} steps are assessed based on performance metrics of the model (following the requirements' analysis). While the metrics are computed on the generated tests at the \textbf{test evaluation} step, we use an independent test set in the \textbf{independent test evaluation}. This independent tests reflect the target distribution of the model and matches the distributions on which the model that is to be deployed. 

Iteratively, the bugs found using the test evaluation can be used to augment the training set 
and trigger a new retraining from scratch. They can also be used to diagnose faulty weights and to directly repair the network, with fine-tuning or with mutation\cite{sohn2022arachne}. 

The model is ready for deployment when the requirements are met on the independent test evaluation. Once the model is deployed, a final evaluation is needed: \textbf{online testing}. During online testing, the performance metrics of the model are evaluated in relation to the whole system, and with the users requests. 


The remaining components of the pipeline in figure \ref{fig:ml_pipeline} are similar to the pipeline proposed by \citet{zhang2020machine}. The scope of our study focuses on the test generation steps and their use in test evaluations and in repair/retrain of the model.

\subsection{Testing requirements}

An ML model can be tested against different requirements 
, i.e., conditions that the model should satisfy when deployed. These requirements are either functional (depend on the model's output predictions) or non-functional (like privacy and efficiency). We focus on functional properties that can be assessed by checking the model predictions against expected values. Hence, we extend the requirements proposed by \citet{zhang2020machine} by considering:

\paragraph{Correctness} refers to the ability of the model to behave satisfactory on unseen inputs with identical distribution to its training data. It measures whether the learning process achieved the best bias-variance trade-off. It is generally measured with metrics like accuracy, F1 score, AUC, ... on the test set.



\paragraph{Generalization.} Contrary to \citet{zhang2020machine}, we distinguish between the performance of the model on the same distribution as the training data, and we refer it as correctness, from the performances of the model on an unknown distribution that is assumed to be a slightly different distribution than the training data's distribution. The ability of a model to generalize to unseen distributions is fundamental for a reliable deployment of an ML model. Indeed, real world online data may slightly differ from the training data.


\paragraph{Robustness.} Robustness in machine learning differ from the definition of robustness by IEEE\cite{159342}\footnote{The degree to which a system or component can function correctly in the presence of invalid inputs or stressful environmental conditions.} in the way that robustness in ML covers both correct and incorrect inputs. Robustness measures how much the behaviour of the ML model will be affected when small perturbations are introduced in the inputs. The perturbations can be adversarial (i.e., designed to cause the misbehaviour of the model) or natural (designed to mimic natural perturbations). 



Because these properties are not always positively correlated,\footnote{Under certain conditions, there exists a trade off between correctness and robustness.} different test sets may be needed to assess different properties. Thus, the adequacy of a test set --- by extension, of an ML testing method -- depends on the considered testing properties. 





\subsection{Related work} Our work takes inspiration from three seminal papers. \citet{zhang2020machine} pioneered the study of ML testing landscape and pipelines, however, studying the evaluation practices of these methods and providing recommendation was out of scope. \citet{Carlini2019OnEA} proposed a white paper on the good practices of evaluating adversarial robustness. Our work and theirs intersect on multiple pitfalls and recommendations, but our contribution goes further with an assessment of the prevalence and the demonstration of their impacts on the results of multiple ML testing frameworks. In addition, our study covers three performance requirements and not only robustness. \citet{arp2020and} provided a critical study on the pitfalls of ML testing through the lenses of the Security community. While we share the same motivation and identified two common pitfalls, we do not assess the same papers, nor evaluate the same performance requirements of ML models.  
\section{Methodology}
\label{sec:methodology}

We introduce the paper collection protocol, an initial analysis of the collected papers, and positions the papers within the ML pipeline and requirements introduced in Section \ref{sec:scope}. The replication package in section \ref{sec:discussion} lists the collected papers and the filtering process.

\paragraph{Data collection.}
The scope of our work is test generation and evaluation following the ML requirements we identified above. We collect the papers starting with a keyword search on the main scientific repositories: Google Scholar and ArXiv. We also used the same keywords on GitHub as many research papers are providing replication packages and source code there. The keywords we used for this search are: \textit{\{(machine learning / ml) (test / testing)\},  \{(deep learning / dl) (test / testing)\}, \{(machine learning / ml) test input generation\}, \{(machine learning / ml) fuzzing\}.}

We complement this keyword-based search with snowballing. We recursively add to our list all the papers whose techniques are compared to in the papers under evaluation, and include also the relevant papers from the related work sections of the analyzed papers. We filter the papers for duplicates, journal extensions, surveys and only include papers from 2018 to 2022. 
All in all, our studies identified 86 papers.

\paragraph{Data distribution.}

Most of the collected papers (70.1\%) have been published in conferences while journal papers account for 17.2\% of the collected papers.
 The non-peer reviewed papers represent 11 papers.
The collected papers are mostly from the Software Engineering community (68 papers), then the Machine Learning community (13 papers).
The communities have been manually annotated using the venues categorization: ICSE, ISSTA, FSE, TSE are for instance categorized as Software Engineering (SE), while NeurIPs, ICML, CVPR, Pattern Recognition, are categorized as machine learning. The ``Other'' category includes papers from the security community, the media and communication, or specialized venues (transportation, smart grids, ...). When the publication is only available on ArXiv 
we use the ArXiV labels to infer its community.



\paragraph{Requirements per pipeline step.}

Among the collected papers, we restrict our studies to papers that have been published in CORE A/A* software engineering venues. 

This additional filtering resulted in 30 publications. We organize the papers in Table \ref{tab:objectives_properties} according to the requirements they aim to improve and the pipeline steps they cover.


Among the 30 filtered papers, 21 do not improve any ML requirement, even if 19 of them aim to generate new test inputs. These inputs are evaluated by their sole ability to increase the coverage of the models, and not by the ability of the inputs to improve the model's requirements.

Among the 9 papers that improve the model's requirements, 5 improve its correctness and 4 improve its robustness. None of the papers have assessed the improvements of the model's generalization performance using the generated inputs.

\begin{table}[]
    \centering
    \small
    \caption{Distribution of analyzed papers by pipeline and requirement}
    \begin{tabular}{l|c|c|c|c|c|}
    \toprule
       Requirement $\downarrow$ & All & Generation  & Priorization & Detection & Oracle \\\midrule
       All              & 30 & 26 & 7 & 6& 0\\    
       None     & 21 & 19 & 2 & 4  & 0\\
       Correctness      & 4 & 3 & 2 & 1& 0\\
       Robustness       & 5 & 4 & 3 & 1 & 0\\
       Generalization   & 0 & 0 & 0 &0 &0 \\
         \bottomrule
    \end{tabular}
    \label{tab:objectives_properties}
\end{table}
\section{Hazards in ML Testing} 
\label{sec:hazards}

Despite the increasing maturity of ML testing research, the proper evaluation of these methods remains non-trivial and prone to different hazards which may bias or invalidate experimental conclusions. Although some of these hazards seem obvious, their practical consideration present subtleties that researchers and users may easily overlook. Our goal is to raise awareness of these issues within the software engineering community, which has a much shorter history dealing with ML models than the core machine learning research and may, therefore, not be acquainted with all the best practices and SoTA that the latter community has contributed to shape.

We report on 10 critical hazards whose manifestation, if not handled properly, can drastically alter experimental conclusions. Later on, we discuss additional factors that may or may not increase confidence in the obtained conclusions, though to a lesser extent than the aforementioned critical hazards. Table \ref{tab:hazard_requirement} summarizes the 10 hazards and the testing requirements for which they are relevant.

\begin{table}[]
\small
\caption{\label{tab:hazard_requirement}In green, the hazards that impact each of the testing requirements}
\setlength\tabcolsep{4pt}
\begin{tabular}{|l|l|l|l|l|l|l|l|l|l|l|}
\hline
               & H1                       & H2                       & H3                       & H4                       & H5                       & H6                       & H7                       & H8                       & H9                       & H10                      \\ \hline
Correctness    & \cellcolor[HTML]{32CB00} &                          & \cellcolor[HTML]{32CB00} & \cellcolor[HTML]{32CB00} & \cellcolor[HTML]{32CB00} & \cellcolor[HTML]{32CB00} & \cellcolor[HTML]{FFFFFF} & \cellcolor[HTML]{32CB00} & \cellcolor[HTML]{32CB00} &                          \\ \hline
Robustness     & \cellcolor[HTML]{32CB00} & \cellcolor[HTML]{32CB00} & \cellcolor[HTML]{32CB00} & \cellcolor[HTML]{32CB00} & \cellcolor[HTML]{32CB00} & \cellcolor[HTML]{32CB00} & \cellcolor[HTML]{32CB00} & \cellcolor[HTML]{32CB00} & \cellcolor[HTML]{32CB00} & \cellcolor[HTML]{32CB00} \\ \hline
Generalization & \cellcolor[HTML]{32CB00} &                          & \cellcolor[HTML]{32CB00} & \cellcolor[HTML]{32CB00} & \cellcolor[HTML]{32CB00} &                          & \cellcolor[HTML]{32CB00} & \cellcolor[HTML]{32CB00} & \cellcolor[HTML]{32CB00} &                          \\ \hline

\end{tabular}
\end{table}

    
    

For each hazard and paper, we apply a 5-scale evaluation. When the hazard is inapplicable or irrelevant to the testing requirement targeted by the method, we mark the hazard as not applicable (\emph{N/A}). We also mark \emph{N/A} the cases when the criteria was not available at the publication of the paper (for example robust models). This is for instance the case for six of the evaluated papers where no rigorous robustification mechanism was available at publication. 

\begin{figure*}
    \centering
    \includegraphics[width=\textwidth]{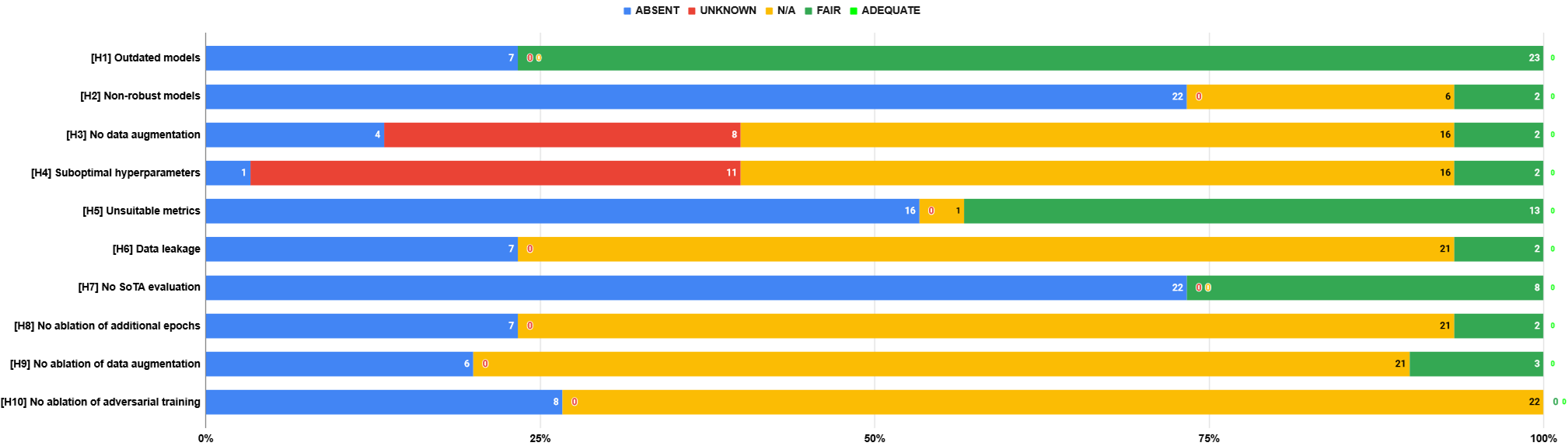}
    \caption{Prevalence of each hazard across the reviewed papers}
    \label{fig:hazards_prevalence}
\end{figure*}

Then we label as \emph{Absent} when the paper has not considered a relevant hazard, \emph{Fair} when the paper has partially addressed the hazard, and \emph{Adequate} when the hazard is addressed following the state of the art. Finally, we use the label \emph{Unknown} for cases where it is unclear (from the paper and its replication package, when available) whether the paper addresses the hazard. Figure \ref{fig:hazards_prevalence} summarizes the prevalence of the hazards in the studied papers.

Each subsection below describes and reports on hazards related to a distinct phase of the ML testing workflow. We describe each hazard, indicate to which testing properties it is relevant, assess its prevalence in the surveyed papers and provide recommendations for future research to properly address it. 

\subsection{Model selection}

Model selection hazards are relevant to experimental protocols involving the use of already available (``pretrained'') models. 

\hazard{H1}{Outdated models}{all properties}%
{The state-of-the-art in machine learning evolves fast with better models being released on a regular basis. The evaluation of ML testing methods on outdated models may introduce a false sense of effectiveness and ultimately produce models that perform worse than the state-of-the-art (even non-improved by testing).}%
{figures/minis/placeholder}\\

\noindent \textbf{Assessment.} Seven over 30 of the investigated papers have used in their evaluation models already outdated at the time of publication (\emph{Absent}). The remaining 23 papers have relied on commonly used models (\emph{Fair}), though none of them included what was the top of the state of the art at publication time.

\noindent \textbf{Recommendations.} While one can appreciate the intense pace at which new ML models are released, there exist established benchmarks and model repositories with state-of-the-art pretrained models (\citet{hendrycks2019robustness} for distribution shifts for example). ML frameworks like Tensorflow and Pytorch also regularly update their set of built-in models, which span across multiple standard datasets. When the target task/datasets are uncommon and pretrained models not available and effective pre-trained models are not available, we recommend to start from pre-trained models trained on similar task and to fine-tune them to the target task.

\hazard{H2}{Lack of robustified models}{robustness}%
{Robustness should be studied on already robustified models \cite{Carlini2019OnEA}. Evaluations with non-robust models lead to inflated success rates of adversarial attacks, thus, robustification methods can improve non-robust models more than robust ones.}%
{figures/minis/placeholder}\\

\noindent \textbf{Assessment:} 22 papers evaluated robustness improvements only on non-robust models (\emph{Absent}). Only two papers consider models equipped with basic/weak robustification mechanisms (\emph{Fair}). None of them used any state-of-the-art robust models at the time of publication. The remaining six were published at a time when no robust models were publicly available.

\noindent \textbf{Recommendations.} There exist repositories and benchmarks for robust models. Notably, Robustbench \cite{croce2020robustbench} includes a large variety of robust models and an API to use them. Popular robustness framework like ART \cite{art2018} and TorchAttacks \cite{kim2020torchattacks} also provide APIs to robustify ML models. Some robustification mechanisms are more reliable than others \cite{Carlini2019OnEA,tramer2020adaptive}. Training with worst-case adversarial examples (adversarial training) remains the most reliable way to obtain a robust model \cite{pgd_madry2019deep}. 



\subsection{Model Training}

Hazards related to model training are applicable to testing methods evaluated on models trained from scratch or fine-tuned for the target task, as opposed to pretrained models. Among the considered papers, 16 used pretrained models only (\emph{N/A}).

\hazard{H3}{No data augmentation}{all properties}%
{Data augmentation is among the most effective ways to optimize all three model properties\cite{carmon2019unlabeled, hendrycks2021many}. Because ML testing methods improve models by generating new inputs, it is essential to evaluate them on data-augmented models to demonstrate the added value of the methods. Failing to do so may result in giving merits to the way the testing methods produce the data, rather than to the simple fact that there are more training data.}%
{figures/minis/placeholder}\\

\noindent \textbf{Assessment:} 12 papers train models without data augmentation (\emph{Absent}) or do not mention doing so (\emph{Unknown}). No paper applied advanced data augmentation techniques (\emph{Adequate}), whereas only two papers used basic ones (\emph{Fair}). 

\noindent \textbf{Recommendations.} We consider basic data augmentations as a minimal requirement, while advanced data augmentations are recommended to stress test the proposed approaches. Basic data augmentations were shown to already yield significant improvements \cite{cubuk2020randaugment}. For computer vision tasks, common data augmentations include image transformations (rotation, cropping, scaling) and their random combinations. 
More advanced augmentation in computer vision include image mixing techniques (e.g. Mixup\cite{zhang2017mixup}, Cutmix\cite{yun2019cutmix}, Augmix\cite{hendrycks2019augmix}), 2D and 3D common corruptions \citet{cubuk2018autoaugment,kar20223d}, unsupervised learning augmentations with unlabeled data\cite{carmon2019unlabeled} and self-supervised augmentations\cite{chen2020adversarial}, or generative model augmentations\cite{gowal2021improving}. 

\hazard{H4}{Suboptimal hyperparameters}{all properties}%
{The training of ML models requires setting multiple hyperparameters like learning rate, loss and architecture optimizations. Suboptimal parameters lead to suboptimal models, which easy bug revelation, i.e., making artificial room for improvement, leading to overestimated test effectiveness. Similarly, the approaches for ML testing often include hyper-parameters, e.g., thresholds, which optimization (or lack off) leads to unreliable results.}%
{figures/minis/placeholder} \\

\noindent \textbf{Assessment.} 12 papers do not optimize hyper-parameters (\emph{Absent}) or do not mention doing so (\emph{Unknown}). Just like data augmentation, only two papers use one set of default hyper-parameter values (\emph{Fair}) and none applied state-of-the-art optimizations available at the time of publications (\emph{Adequate}).

\noindent \textbf{Recommendations.} When the paper trains custom models, a minimal requirement is to use common training parameters. 
These parameters include training parameters such as learning rate, loss optimizer, and architecture optimizations like split batch normalization. 
A recommended approach is for the paper to test out several hyper-parameters on the specific task, model and dataset using optimized hyper-parameter search libraries such as Optuna\cite{akiba2019optuna}.

\subsection{Evaluation}

\hazard{H5}{Unsuitable metrics}{all properties}%
{Measuring the effectiveness of a testing method using inappropriate metrics may result in fallacious conclusions. }
{figures/minis/placeholder}

\hazard{H5}{}{continue}%
{Model performance should be measured with standards metrics (accuracy, AUC, MSE, etc.) or domain-specific ones. Depending on the requirement (correctness, robustness, generalization), one should also accordingly measure the similarity and distribution of generated inputs.}
{figures/minis/placeholder}\\

\noindent \textbf{Assessment.} 16 papers rely on spurious metrics to evaluate testing methods (\emph{Absent}). For instance, the number of generated bugs (misclassified examples) is not a conclusive metrics because one can trivially generate many $\epsilon$-close examples from a revealed bug. Other metrics like neuron coverage were proposed and rapidly dismissed as a relevant evaluation metric\cite{harel2020neuron}. 13 over the 30 papers use common model evaluation metrics (\emph{Fair}) whereas none also evaluates the validity of the labels or checks distribution of the generated inputs (\emph{Adequate}). 

\noindent \textbf{Recommendations.} Our key recommendation is to systematically measure the value of a testing method as its contribution to improving model performance (using the established and relevant metrics). The chosen metrics should also be consistent across evaluations. When new metrics are provided in replacement,  demonstrating their correlation to well established metrics is not sufficient. The correlations can be spurious or alter the order of magnitude of the achieved improvements. For robustness, we recommend measuring it via the empirical success rate achieved by state-of-the-art attacks (e.g., AutoAttack\cite{autoattack}), using appropriate the recommended parameters proposed by \citet{croce2020robustbench}. One step further, we also recommend using certifiable methods (e.g. CLEVER \cite{clever_weng2018evaluating} or randomized smoothing\cite{smoothing_cohen2019certified}). 

In addition, when generating new inputs, additional metrics such as the SSIM\cite{hore2010image}, LPIPS\cite{zhang2018perceptual} or the $L_p$ distance to the original seeds are needed if the labels are assumed unchanged, and ensuring that the new inputs belong to the same distribution as the original seeds is needed when assessing the correctness of the model.


\hazard{H6}{Data leakage}{correctness and robustness}{Testing methods typically generate tests from existing inputs (``seeds''). Using the same seeds to evaluate an improved model is a form of data leakage and may yield overestimated effectiveness.}%
{figures/minis/placeholder} \\

\noindent\textbf{Assessment.} seven papers evaluate retrained models on the same seeds used to produce the retraining data (\emph{Absent}). 21 do not retrain models at all (\emph{N/A}). two papers keep part of the test set as an independent evaluation set (\emph{Fair}). The ideal practice of generating the retraining data from the training set and using the test set only for evaluation (\emph{Adequate}) is not followed by any paper.

\noindent\textbf{Recommendations.} An independent test set should be held out in order to measure model performance after retraining. This set can be for measuring both correctness, generalization and robustness (via the application of adversarial attacks or other test generation methods on this independent set).

\hazard{H7}{No SoTA evaluation}{robustness and generalization}%
{A rigorous evaluation should rely on state-of-the-art techniques, especially in robustness evaluation. The use of simple image transformations and corruptions, weak attacks like FGSM \cite{fgsm_goodfellow2015explaining}, or strong attacks, like PGD4 \cite{pgd_madry2019deep}, with small budget leads to overestimated robustness and generalization.}
{figures/minis/placeholder}\\

\noindent\textbf{Assessment.} None of the papers evaluated their approaches using SoTA evaluation techniques, and 22 papers did not evaluate their approaches using common baselines.

\noindent\textbf{Recommendations.} We recommend that test input generation approaches compare to the strongest attacks, using both gradient based (parameter-free like AutoPGD\cite{autopgd_croce2020reliable} and black box approaches like SparseRS \cite{croce2022sparse}). Ensemble-based attacks, while costly could also be considered for exhaustive evaluation (AutoAttack \cite{autoattack}). Test selection and prioritization papers should not only consider using popular metrics from the Software Engineering community like DeepGini\cite{feng2020deepgini}, Entropy\cite{shannon1948mathematical}, and surprise \cite{kim2018guiding}, but extend their evaluation to the SoTA from the ML community\cite{akhtar2021advances}.

\subsection{Repair/Retraining}

\hazard{H8}{Additional epochs}{all properties}%
{\citet{swa_izmailov2018averaging} demonstrated that Stochastic Weight Averaging (SWA) and cyclic learning rate improve the correctness and generalization of neural networks. Fine-tuning the models for additional epochs with the same training data and optimal learning rate optimizers can already improve the performances.}
{figures/minis/placeholder}\\

\noindent \textbf{Assessment.} While 21 papers do not fine-tune the models (\emph{N/A}), seven papers fine-tune them with new test inputs without assessing if the improved performances can be explained by the additional training budget (\emph{Absent}). Only two papers evaluate the additional fine-tuning baseline, and none of the assessed the best fine tuning practices such as SWA and cyclic learning rates.







\noindent\textbf{Recommendations.} Fine-tuning models for a few epochs using the same training data can be enough and should be a sanity check to assess any improvement achieved by test input generation techniques. More expensive fine-tuning approaches like SWA should also be considered for an extensive evaluation of the contribution of fine-tuning.

\hazard{H9}{No data-augmented repair/retraining}{all properties}%
{Similarly to evaluating the impact of data augmentation on the initial models, any process that involves fine-tuning the model should consider the impact of data augmentation in replacement (and in combination) of the proposed augmented inputs. This is important to ensure that all testing and retraining methods are applied with the full potential available. Ultimately, what matters is to produce the best model combining available methods.} 
{figures/minis/placeholder}\\

\noindent \textbf{Assessment.} six papers do not use data augmentation at all during retraining \emph{Absent}. Only three use basic data augmentation (like image transformations) and none use state-of-the-art data augmentation methods jointly with the testing/retraining approach \emph{Fair}. None combined their approach with advanced data augmentation.

\noindent \textbf{Recommendation.} Our recommendations follow those given for \textbf{H3} and to apply data augmentation also during the retraining phase. This enables researchers to showcase the benefits of their retraining method when combined with state-of-the-art data augmentation methods. To fully address this hazardn we recommend the use of a combination of advanced data augmentations including unlabeled data\cite{carmon2019unlabeled} and generative models augmentation\cite{gowal2021improving}. 

\hazard{H10}{No adversarial training}{robustness}%
{Madry's adversarial training \cite{pgd_madry2019deep} is a min-max optimization training where each batch of the training includes both an original training example and its associated adversarial example generated with a PGD attack. 
It is important both the assess how adversarial training would impact the robustness of the evaluated models and how adversarial training could be combined with any proposed input generation approach.}
{figures/minis/placeholder}\\

\noindent \textbf{Assessment.} 22 papers did not have to address this hazard (\emph{N/A}): 21 did not target robustness, and one was published before adversarial training was demonstrated to be effective. None of the remaining papers evaluated the impact of adversarial training.

\noindent \textbf{Recommendation.} A fair evaluation would require comparing any approach to basic Madry adversarial training. An adequate evaluation entails to also compare how the proposed approach interacts with adversarial training, and assessing advanced variants of adversarial training, such as FREE \cite{shafahi2019free}, TRADES \cite{zhang2019trades}, MaxUp \cite{gong2020maxup}, and GAT\cite{ghamizi2023gat}.



\section{Empirical Analysis of the hazards}
\label{sec:impacts}

\subsection{Experimental protocol}

\paragraph{Testing methods (subjects)} We consider six popular ML testing methods whose implementation is publicly available: DeepXplore \cite{pei2017deepxplore}, SADL \cite{kim2018guiding}, ADAPT \cite{adapt_lee2020effective}, DeepSearch \cite{zhang2020deepsearch}, DeepGini \cite{feng2020deepgini}, and RobOT \cite{wang2021robot}. These methods rely on one or multiple metrics to select test inputs: neuron coverage (used by DeepXplore and ADAPT), surprise adequacy -- LSA/DSA (SADL and ADAPT), First-order Loss (FoL) (DeepSearch and RobOT), output probabilities (DeepGini). Because our goal is not to finger-point specific methods/papers but rather to provide a fine-grained analysis of how the hazards can impact testing methods and actionable recommendations for future research, we evaluate not the methods altogether but the metrics they rely on independently.

We thus compute each metric with their default parameters: a threshold of 0.5 for neuron coverage, a bucket size of 100 features for LSA, a badge size of 10 for DSA. Both LSA and DSA are measured on the penultimate layer (as in the original paper) and the neuron coverage is measured on all the hidden layers (i.e. except the logits layer). We evaluate the impact of each hazard to each method, considering all testing objectives: clean accuracy (correctness), adversarial accuracy (robustness), and accuracy on corruptions (generalization). 

\paragraph{Models and datasets}


Following the Robustbench benchmark, we pick 6 models pretrained for CIFAR-10 with varying architectures, training protocol and performances. Five of these models have been robustified using state-of-the-art data augmentation and adversarial training techniques on the CIFAR-10 dataset. \textbf{Standard} \cite{croce2020robustbench} is a WideResNet-28-10 model trained with standard training and common data augmentation. \textbf{Engstrom} \cite{engstrom_robustness} is a Resnet-50 model robustified with adversarial training and common data augmentation. \textbf{Carmon} \cite{carmon2019unlabeled} is ``Standard'' with adversarial training using unlabeled data. \textbf{Rebuffi} \cite{rebuffi2021fixing} is a WideResNet-70-16 adversarially trained with unlabeled data, stochastic weight averaging, and advanced data-augmentation (CutMix \cite{yun2019cutmix}). \textbf{Gowal28} and \textbf{Gowal70} \cite{gowal2021improving} are, respectively, a WideResNet-28-10 and WideResNet-70-16 improved with data augmentation.

\paragraph{Evaluation sets and model performance metrics.} We measure model accuracy on original test examples (for correctness), on corrupted examples (for generalization) and on adversarial examples (for robustness). We thus use three evaluation sets. For correctness, we pick 1000 test examples from the original CIFAR-10 dataset. For robustness, we generate adversarial examples from the same set of 1000 original examples (one adversarial per original example) with AutoAttack-8, the best method recommended by Robustbench. For generalization, we take 1000 examples from CIFAR-10C \cite{hendrycks2019robustness}, a dataset that contain CIFAR-10 images with common corruption (rotation, brightness change, etc.). This dataset is commonly used to emulate distribution shifts within CIFAR-10.

    

\paragraph{Evaluation steps}
We divide our evaluation along the two objectives of ML testing: test selection (the problem of selecting misclassified inputs \cite{ma2019test}) and model repair. In both steps, we consider the selection/use of original test examples, adversarial examples, and corrupted (shifted) examples. We evaluate the impact of each hazard in the first step where it has an impact, i.e. hazards H1--H5 have already an effect in the test selection step whereas H6--H10 intervene during model repair. As our results will reveal, the test selection step already demonstrates the negative impact of hazards H1--H5. Note that our results for the model repair step also contain the impact of those hazards studied in the test selection step, although we do not discuss this impact in detail for the sake of avoiding redundancies.

\paragraph{Experimental settings for test selection (Step 1)}

In this step, we aim to measure the capability of the studied metric to select a set of inputs likely to be misclassified. Because most metrics are used to select \emph{sets} rather than individual examples (e.g. neuron coverage aims to maximize the number of neurons activated at least one by a set of examples) -- while the metrics applicable to single examples can be straightforwardly applied to a set instead -- we separate each of our evaluation sets into 40 random buckets and measure the capability of the metric to select the set containing the highest number of misclassified examples. We measure this capability by computing the Spearman correlation between (1) the metric value given to each bucket and (2) model accuracy on the bucket. A metric performs better as it has a stronger negative correlation to accuracy. Note that the FoL metric can only be computed between an original input and a locally perturbed input, thus it can only be measured against adversarial accuracy.

\paragraph{Experimental setting for model repair (Step 2)}

We aim to measure the capability of each testing metric to repair the models using the examples the metric selects. We repair via fine-tuning, i.e. training the models for additional epochs, which is a popular way to improve models using selected or generated data (as opposed to retraining from scratch, which is computationally more expensive). We then study two settings: standard fine tuning (using non-adversarial data only) and Madry adversarial fine-tuning (using non-adversarial and their associated adversarial data). In each case, we measure the accuracy of the fine-tuned model on the three evaluation sets (clean original examples, AutoAttack examples, corrupted examples).

For standard fine-tuning, we fine-tune each pre-trained model for 40 additional epochs using a One Cycle learning rate scheduler (min=0.005, max=0.01). During fine-tuning we use 50\% of the original training set and we replace the other with 10\% inputs of its inputs (seeds) selected by the metric. We ensure that only one new example is generated from one seed. We compare the fine-tuned models with three baselines: (1) the model fine-tuned using the full original training set (without replacement), (2) the model fine-tuned with half of the original training set and PGD adversarial examples produced by the other half and (3) the model fine-tuned with half of the original training set, and the other half replaced with augmented data (using color and geometric augmentations from AutoAugment \cite{cubuk2018autoaugment}). It is to be noted that, because we use data augmentation, we make sure that all fine-tuned models have been trained with the same amount of different data. 

For adversarial fine-tuning, we repeat the same experiments except that we do not fine-tune models with the selected seeds but with the selected seeds \textbf{and} with adversarial examples generated from these seeds. To generate these adversarial training examples, we use PGD with $\epsilon=8/255$ in $\ell_\infty$ norm, and 10 steps.

\begin{table}[t]
    \centering
    \small
\caption{\label{tab:metric_retrain_adversarial_leakage}Data leakage: adversarial fine-tuning using each metric.}
\begin{tabular}{l|rrr|rrr}
\toprule
 & \multicolumn{3}{l}{Correctness} & \multicolumn{3}{l}{Robustness (AutoAttack)} \\
Model &  DeepGini & LSA &  NC &  DeepGini &  LSA &  NC \\
\midrule
Standard &                 70.3 &            69.7 &           81.7 &                26.6 &            8.3 &          27.3 \\
Engstrom   &                 93.2 &            83.8 &           91.5 &                58.6 &           55.9 &          57.4 \\
Carmon     &                 97.2 &            90.7 &           95.2 &                67.8 &           54.7 &          62.7 \\
Gowal28    &                 99.7 &            99.5 &           99.5 &                96.2 &           92.4 &          87.4 \\
\bottomrule
\end{tabular}

\end{table}

\subsection{Impact of SoTA models and data augmentations (test selection)}

We summarize our results in Table \ref{tab:metric_correlation}. We see that none of the metrics are correlated with the generalization performance. Neuron coverage is not correlated with the clean performance of models trained by Gowal et al. that rely on large-scale data augmentations. 
LSA is not correlated with any performance metric for the 2 WideResnet28 adversarially trained models (Carmon and Gowal28).  DeepGini is not correlated with any performance metric for the model proposed by Carmon et al. FoL is not correlated with the adversarial performance of models trained by Gowal et al. that rely on large-scale data augmentations. 

The lack of correlations in some settings suggests that the effectiveness of selection metrics does not generalize. Proper assessment should therefore not focus on these metrics but rather on the relevant model performance metrics (e.g. the accuracy on original test, on adversarial examples, and on distribution shifted data).

\pitfallenv{H1, H2, H3, H5}{SoTA architectures, augmentations, and robustification}%
{\label{pit:sota} The metrics used to test and generate test data are ineffective for some architectures, training protocols and testing requirements (esp. correctness and generalization).}%

\begin{table*}[]
\caption{\label{tab:metric_correlation}Correlations between testing metrics and each requirement performance. In bold the correlations with p-value < 0.05. The accuracy is over the full set. NC: Neuron Coverage, LSA and SDA are Surprise Adequacy and FoL is First-order loss of RoBOT.}
\small
\scalebox{1}{
\begin{tabular}{l|l|rrrrrl}
\toprule
{Models}                                   & {Evaluation set}      & {Accuracy} & {NC } & {LSA}            & {DSA} & {DeepGini} & FoL                                  \\ \cline{1-8}
Standard                                        & Clean                            & 94.78                               & 0.083                                & \textbf{-0.768}                     & \textbf{-0.787}         & 0.142                        & - \\
(non-robust)                                      & AutoAttack 8/255                        & 0.00                                & {-}                & {-}               & {-}   & {-}        & -                                            \\ 
             & {Corruptions} & {57.50}          & {0.195}           & {0.316}          & -0.129                  & 0.089                        & -                                            \\  \midrule
Engstrom     & {Clean}       & {87.03}          & {\textbf{0.688}}  & {\textbf{0.829}} & \textbf{-0.816}         & \textbf{0.746}               & -                                            \\ 
                                      & AutoAttack 8/255                        & 49.25                               & \textbf{0.927}                       & \textbf{0.906}                      & \textbf{-0.936}         & \textbf{0.967}               & {-0.14}                    \\  & Corruptions                      & 71.60                               & \textbf{0.454}                       & 0.183                               & -0.211                  & -0.09                        & -                                            \\ 
\midrule
{Carmon}                                                                     & Clean                            & 89.69                               & \textbf{0.341}                       & -0.093                              & \textbf{-0.878}         & 0.199                        & -                                            \\
                                                               & AutoAttack 8/255                        & 59.53                               & \textbf{0.794}                       & 0.184                               & \textbf{-0.954}         & 0.348                        & {\textbf{0.408}}           \\                    & Corruptions                      & 69.20                               & 0.024                                & 0.052                               & -0.208                  & -0.165                       & -                                            \\
\midrule
{Rebuffi}                                                                 & Clean                            & 92.23                               & \textbf{0.469}                       & \textbf{0.914}                      & \textbf{-0.852}         & \textbf{0.752}               & -                                            \\
                                                               & AutoAttack 8/255                        & 66.56                               & \textbf{0.928}                       & \textbf{-0.963}                     & \textbf{-0.938}         & \textbf{0.862}               & {\textbf{0.474}}           \\       & Corruptions                      & 72.60                               & 0.296                                & -0.197                              & -0.395         & -0.04                        & -                                            \\
\midrule
{Gowal70}                                                                & Clean                            & 88.74                               & 0.251                                & \textbf{-0.83}                               & \textbf{-0.89}          & \textbf{0.472}               & -                                            \\
                                                               & AutoAttack 8/255                        & 66.10                               & \textbf{0.823}                       & \textbf{-0.878}                     & \textbf{-0.953}         & \textbf{0.526}               & {0.267}                    \\          & Corruptions                      & 71.90                               & -0.216                               & -0.017                              & -0.037                  & -0.176                       & -                                            \\
\midrule
{Gowal28}                                                               & Clean                            & 87.50                               & 0.052                                & -0.11                               & \textbf{-0.928}         & \textbf{0.713}               & -                                            \\
                                                               & AutoAttack 8/255                        & 63.38                               & \textbf{0.792}                       & 0.277                               & \textbf{-0.959}         & \textbf{0.858}               & {0.271}                    \\           & Corruptions                      & 69.60                               & 0.206                                & -0.165                              & -0.28                   & -0.05                        & -  \\
\bottomrule
\end{tabular}
}
\end{table*}

\subsection{Impact of hyper-parameters (test selection)}

Neuron coverage, LSA and DSA rely on multiple parameters such as the activation threshold and the boundaries. For these two metrics, we evaluate variants with different hyper-parameters.

\begin{table}[]
\small
\caption{\label{tab:metric_correlation_params}Correlations between metrics and model performance when changing the hyper-parameters. In bold the correlations with p-value < 0.05.}
\begin{tabular}{l|l|rrrrr}
\toprule
Models                 & Evaluation set      & {NC 0.25} & {NC 0.5} & {NC 0.75} & {LSA 300} & {LSA 100} \\ \midrule
Standard             & Clean       & 0.111                       & 0.083                      & -0.007                      & \textbf{-0.782}             & \textbf{-0.768}             \\
(non-robust)                       & AutoAttack 8/255   & -       & -      & -       & -       & -       \\
                       & Corruptions & 0.254                       & 0.195                      & 0.259                       & \textbf{-0.329}             & 0.316                       \\ \midrule
Engstrom & Clean       & \textbf{0.662}              & \textbf{0.688}             & \textbf{0.718}              & \textbf{0.629}              & \textbf{0.829}              \\
                       & AutoAttack 8/255   & \textbf{0.886}              & \textbf{0.927}             & \textbf{0.933}              & \textbf{0.878}              & \textbf{0.906}              \\
                       & Corruptions & \textbf{0.56}               & \textbf{0.454}             & \textbf{0.356}              & 0.222                       & 0.183                       \\ \midrule
Carmon    & Clean       & \textbf{0.548}              & \textbf{0.341}             & 0.139                       & 0.091                       & -0.093                      \\
                       & AutoAttack 8/255   & \textbf{0.909}              & \textbf{0.794}             & 0.56                        & \textbf{0.508}              & 0.184                       \\
                       & Corruptions & 0.131                       & 0.024                      & -0.069                      & -0.088                      & 0.052                       \\ \midrule
Rebuffi            & Clean       & -0.008                      & \textbf{0.469}             & \textbf{0.492}              & \textbf{-0.838}             & \textbf{0.914}              \\
                       & AutoAttack 8/255   & \textbf{0.59}               & \textbf{0.928}             & \textbf{0.799}              & \textbf{-0.967}             & \textbf{-0.963}             \\
                       & Corruptions & \textbf{-0.457}             & 0.296                      & \textbf{0.462}              & -0.198                      & -0.197                      \\ \midrule
Gowal70   & Clean       & -0.202                      & 0.251                      & \textbf{0.394}              & \textbf{-0.857}             & -0.83                       \\
                       & AutoAttack 8/255   & \textbf{0.667}              & \textbf{0.823}             & \textbf{0.799}              & \textbf{-0.888}             & \textbf{-0.878}             \\
                       & Corruptions & \textbf{-0.437}             & -0.216                     & 0.127                       & -0.259                      & -0.017                      \\ \midrule
Gowal28   & Clean       & \textbf{-0.488}             & 0.052                      & 0.196                       & -0.092                      & -0.11                       \\
                       & AutoAttack 8/255   & \textbf{-0.692}             & \textbf{0.792}             & \textbf{0.798}              & \textbf{0.585}              & 0.277                       \\
                       & Corruptions & \textbf{-0.649}             & 0.206                      & \textbf{0.425}              & -0.109                      & -0.165                      \\ 
\bottomrule
\end{tabular}

\end{table}

Depending on the selected threshold, neuron coverage can be positively or negatively correlated with the generalization performance for the two WideResnet28 models. 
There is no clear trend on what is the best threshold to use across the models and performance metrics. Even with increased number of features to 300, LSA remains uncorrelated for half of the combinations evaluated.

\pitfallenv{H4}{Hyper-parameters tuning}%
{\label{pit:sota} To be effective, metrics used to test and generate new test examples require careful optimization and tuning specifically for the considered settings.}%

\begin{table*}[]
\small
\caption{\label{tab:metric_retrain_standard} Change of performance when standard fine-tuning using each metric. In bold the best performances per row. PGD: PGD attack with $\epsilon=8/255$, AA: AutoAugment}
\scalebox{0.9}{
\begin{tabular}{l|l|ll|llllll}
\toprule
                                         &                          &        Before                   &      Standard                          & \multicolumn{6}{l}{Standard fine-tuning + training set augmented by} \\
 Models               & {Evaluation set} & {fine-tuning} & {fine-tuning} & {PGD}           & {AA}   & {NC} & {LSA}           & {DeepGini} & FoL                \\ \midrule
Standard                                            & Clean                    & \textbf{94.8}             & 94.6                           & {93.0}          &  93.9    & {93.7}            & {89.5}          & {93.1}     & 12.3 \\ 
(non-robust)                                         & AutoAttack 8/255                & 0                         & 2.2                  & {3.1}           & {1.2}           & {1.4}             & \textbf{4.7}           & {1.5}      & 3.6                         \\ 
            & Corruptions              & 57.5                      & 70.4                           & {\textbf{73.9}} & {67.2}          & {62.9}            & {58.1}          & {62.3}     & 10.3                        \\ \midrule
\multirow{3}{*}{Engstrom}                                         & Clean                    & 87.0                      & \textbf{93.5}                           & {90.4}          & 92.3 & {86.0}            & {83.6}          & {88.6}     & { 91.8}                  \\ 
                                         & AutoAttack 8/255                & \textbf{49.3}                      & 30.7                  & {33.2}          & {26.1}          & {42.9}            & {37.6}          & {42.9}     & 8.8                         \\ 
 & Corruptions              & 71.6                      & 78.8                           & \textbf{80.3}          & 78.5 & {70.4}            & {67.0}          & {72.9}     & 74.3                        \\ \midrule
 \multirow{3}{*}{Carmon}                                        & Clean                    & 91.9                      & \textbf{95.0}                           &  94.2    & 94.7 & {89.8}            & {82.9}          & {92.4}     & { 93.7}                  \\ 
                                         & AutoAttack 8/255                & \textbf{59.5}             & 6.6                           & {16.0}          & {8.2}           & {57.2}            & {45.7}          & {40.4}     & 15.3                        \\ 
    & Corruptions              & 69.2                      & 73.3                           &  76.2    & {\textbf{76.6}} & {72.5}            & {62.3}          & {73.5}     & { 76.6}                  \\ \midrule
 \multirow{3}{*}{Rebuffi}                                                & Clean                    & 92.2                      & \textbf{95.7}                           &  94.5    &  95.1    & {84.2}            & {45.5}          & {72.2}     & {95.2}               \\ 
                                         & AutoAttack 8/255                & \textbf{66.6}             & 19.7                           & {23.3}          & {13.1}          & {32.9}            & {13.4}          & {18.0}     & 20.2                        \\ 
    & Corruptions              & 72.6                      & 74.6                           & {72.4}          & \textbf{75.2}          & {63.0}            & {30.9}          & {51.7}     & 74.5                        \\ \midrule
\multirow{3}{*}{Gowal70}                           & Clean                    & 88.7                      & 90.9                           & {90.5}          &  \textbf{91.4}          & {90.6}            & 87.1             & {89.8}     & 9.9                         \\ 
                                         & AutoAttack 8/255                &{66.1}             & 68.0                           &  \textbf{68.5}          & {42.3}          & {67.9}            & 66.0             & {64.9}     & 3.8                         \\ 
   & Corruptions              & 71.9                      & 74.3                           & {73.0}          &  \textbf{76.0}          & {75.0}            & 71.9             & {75.8}     & 12.3                        \\ \midrule
\multirow{3}{*}{Gowal28}             & Clean                    & 87.5                      & 91.6                           & {90.6}          & {\textbf{91.7}} & {86.2}            & {90.1}          & {89.3}     & 89.9                        \\ 
                                         & AutoAttack 8/255                & \textbf{63.4}             & 53.4                           & {58.7}          & {26.3}          & {57.2}            & {59.5}          & {59.1}     & 56.3                        \\ 
   & Corruptions              & 69.6                      & \textbf{75.2}                           & {72.1}          & {73.7}          & {72.5}            & 75.1 & {74.1}     & 74.5                        \\ \bottomrule
\end{tabular}
}
\end{table*}

\begin{table*}[]
\small
\caption{\label{tab:metric_retrain_adversarial}Change of performance when adversarial fine-tuning using each metric. In bold the best performances per row.}
\scalebox{0.9}{
\begin{tabular}{l|l|ll|llllll}
\toprule
                       &                    & Before        & Adversarial   & \multicolumn{5}{l}{Adversarial fine-tuning + training set augmented by} \\
Models                 & Evaluation set             & fine-tuning   & fine-tuning   &AutoAugment          & NC          & LSA       & DeepGini       & FoL        \\\midrule
Standard             & Clean              & \textbf{94.8} & 86.5          & 85.2                 & 75.1        &    65.5       & 68.2           & 9.2        \\
(non-robust)                       & AutoAttack 8/255   & 0             & \textbf{36.6} & 33.1                 & 23.8        &    7.0       & 20.1           & 9.8        \\
                       & PGD 8/255          & 0             & \textbf{85.0}          & 82.3                 & 68.8        &   29.8        & 64.3           & 8.7        \\
                       & FGSM 8/255         & 30.5          & \textbf{68.9}          & 82.1                 & 67.9        &   36.8        & 62.9           & 8.7        \\
                       & Corruptions        & 57.5          & \textbf{71.0} & 67.7                 & 58.9        &      39.4     & 49.6           & 6.0        \\\midrule
Engstrom & Clean              & 87.0          & 89.4          & \textbf{88.0}        & 85.4        &      77.3     & 86.8           & 85.2       \\
                       & AutoAttack 8/255   & 49.3          & \textbf{54.2} & 54.1                 & 47.2        &      46.9     & 47.0           & 43.6       \\
                       & PGD 8/255          & 55.4          & \textbf{61.7}          & 58.9                 & 55.1        &   52.8        & 54.5           & 57.0       \\
                       & FGSM 8/255         & 61.50         & \textbf{65.4}          & 63.7                 & 57.7        &     56.4      & 59.7           & 60.7       \\
                       & Corruptions        & 71.6          & 71.3          & \textbf{71.6}        & 70.1        &     64.6      & 70.3           & 66.8       \\\midrule
Carmon    & Clean              & 89.7          & \textbf{90.7} & 89.2                 & 88.0        &     81.3      & 87.3           & 89.3       \\
                       & AutoAttack 8/255   & \textbf{59.5} & 57.8          & 54.6                 & 50.4        &    44.7       & 51.0           & 52.6       \\
                       & PGD 8/255          & 63.7          & \textbf{73.1}         & 68.5                 & 65.3        &    58.6       & 64.3           & 64.9       \\
                       & FGSM 8/255         & 70.6          & \textbf{77.3}          & 73.2                 & 68.9        &      63.2     & 69.1           & 70.8       \\
                       & Corruptions        & 69.2          & \textbf{73.3} & 70.0                 & 67.1        &    59.8       & 68.5           & 72.4       \\\midrule
Rebuffi            & Clean              & 92.2          & 90.1          & 88.0                 & 82.3        &  35.5         & 70.8           & 86.0       \\
                       & AutoAttack 8/255   & \textbf{66.6}          & 54.5          & 49.2                 & 50.6        &    21.7       & 44.0           & 48.8       \\
                       & PGD 8/255          & 70.5          & \textbf{75.2}          & 73.7                 & 63.5        &    27.9       & 58.1           & 68.4       \\
                       & FGSM 8/255         & 75.7          & \textbf{77.7}          & 72.2                 & 66.8        &    29.8       & 60.6           & 72.2       \\
                       & Corruptions        & 72.6          & 71.6          & 71.9                 & 76.9        &    21.5       & 54.3           & 71.9       \\\midrule
Gowal70   & Clean              & 88.7          & 89.7          & 86.9                 & 89.8        &   85.3        & 89.4           & 11.3       \\
                       & AutoAttack 8/255   & 66.1          & 64.8          & 52.0                 & 63.8        &   63.3        & \textbf{66.2}           & 5.1        \\
                       & PGD 8/255          & 69.8          & \textbf{73.2}          & 71.9                 & 70.7        &      69.5      & 69.8           & 11.7       \\
                       & FGSM 8/255         & 73.9          & \textbf{76.7}          & 71.9                 & 74.7        &    73.9      & 73.8           & 9.6        \\
                       & Corruptions        & 71.9          & \textbf{73.6}          & 66.6                 & 75.3        & 71.9          & 76.3           & 9.7        \\\midrule
Gowal28   & Clean              & 87.5          & \textbf{90.3} & 89.4                 & 85.9        &     88.0      & 89.0           & 88.3       \\
                       & AutoAttack 8/255   & \textbf{63.4} & 62.2          & 54.3                 & 57.0        &     63.9      & 57.6           & 58.7       \\
                       & PGD 8/255          & 66.3          & \textbf{68.8}          & 71.8                 & 67.2        &      66.9     & 66.4           & 67.4       \\
                       & FGSM 8/255         & 70.4          & \textbf{71.7}          & 72.0                 & 69.0        &      70.8     & 69.1           & 70.2       \\
                       & Corruptions        & 69.6          & 73.3          & \textbf{73.7}        & 72.0        &     73.8      & 72.5           & 70.7      \\
                       
\bottomrule
\end{tabular}
}
\end{table*}

\subsection{Impact of additional epochs and data augmentation (model repair)}
\label{sec:ImpactStandard}

We summarize our evaluation in Table \ref{tab:metric_retrain_standard}. Starting from robust models, fine-tuning them with selected inputs degrades the initial robustness. 
Fine-tuning the undefended model (\textit{Standard}) does not improve robustness regardless of the used selection metrics. This confirms that standard training, when augmented with any test input generation is not sufficient to achieve robust models and suggests that adversarial training remains necessary. 

The best correctness performance is achieved either using standard fine-tuning or with or using AutoAugment fine-tuning. This confirms that additional training is sufficient to improve the correctness of the models.
Meanwhile, the best generalization performance is achieved either by using PGD fine-tuning or by using AutoAugment fine-tuning, confirming that simple data augmentation is sufficient to improve the generalization of the models.

\pitfallenv{H8, H9}{Additional epochs and augmentations}%
{\label{pit:sota} The best improvements in correctness and generalization of the repaired models are achieved via more training and data augmentation, not by specific testing metrics.}%

\subsection{Impact of adversarial training and evaluation attacks (model repair)}
\label{sec:ImpactAdversarial}

Results in Table \ref{tab:metric_retrain_adversarial} confirm that the best improvement in robustness of the repaired models is always achieved by adversarial training. Guiding adversarial training through testing metrics actually decreases the robustness after repair. 

Next, we compare the robustness achieved through each metric against various attacks, FGSM being the weakest and AutoAttack the strongest. Repair using the metrics achieves high robustness against FGSM (e.g. 67.9\% robust accuracy for the undefended models improved with neuron coverage) but is significantly less effective against stronger attacks (e.g. neuron coverage achieves 23.8\% robust accuracy against AutoAttack).

\pitfallenv{H7}{SoTA evaluation}%
{\label{pit:sota} Evaluating robustness against weak attacks gives a false sense of robustness as repaired models perform poorly against stronger attacks.}%

\pitfallenv{H10}{Adversarial training}%
{\label{pit:sota} Classical adversarial training yields better improvements in robustness than the selection metrics.}

\subsection{Impact of data leakage (model repair)}

Comparing the results without data leakage in Table \ref{tab:metric_retrain_adversarial} and with data-leakage in Table \ref{tab:metric_retrain_adversarial_leakage}, we observe for all testing metrics that data leakage leads to overestimating correctness and robustness. For example, the Carmon model has its correctness jump from 88\% to 95.2\% when retrained with neuron-coverage-guided adversarial training, while its robustness jumps from 50.4\% to 62.7\%.

\pitfallenv{H6}{Data leakage}%
{\label{pit:sota} Evaluating models on the same seeds used to generate the new test examples leads to data leakage and this overestimates the correctness and robustness of the models.}%

\section{Discussion}
\label{sec:discussion} 

\paragraph{Additional recommendations.} Our study focuses on 10 critical hazards that, as we have shown, have a significant impact on experimental conclusions. There are additional recommendations we would like to bring up to support more expensive experiments of ML testing methods.

Like most ML testing papers, we focused our study on image classification. We do, however, recommend assessing ML testing techniques on \textbf{other tasks} such as image segmentation, and \textbf{other domains} such as NLP, code, and tabular data. Furthermore, datasets with particular behaviours like temporal drifts or class imbalance should be evaluated with an adequate assessment, e.g. WILDS \cite{beery2020iwildcam}. 

We also want to raise awareness on the relevance of the datasets used. For example, MNIST and the like are not relevant datasets for evaluation as correctness on these dataset is a solved problem and its small size makes it non-relevant to study robustness \cite{croce2020robustbench}.

We contrast our consideration towards repetitions of experiments to ensure statistical significance. Training for multiple batches and epochs is a form of repetition that makes classical experimental repeat unfruitful. We validated this by running our experiments five times and observed a standard deviation of less than 1\%.   

A clear threat model should properly define the considered risks (e.g. whitebox or blackbox adversarial attacks, natural distribution shifts, etc.) and tune the experimental protocol (including chosen baselines) accordingly. Because there is a trade-off between correctness, robustness and generalization \cite{tsiprasSETM19}, a Pareto evaluation should be considered to make sure improving one criterion does not degrade the others. 

In addition to the full method, ablation studies are usually recommended to assess the independent contribution of each component.

Finally, because deep learning is expensive, computation cost should be studied both as a proper evaluation criteria (to balance with achieved improvements) and as an equal comparison ground (fixed budgets). Similarly, adaptive evaluations \cite{tramer2020adaptive} permits to study whether stronger attacks/drifts can break the improvements provided additional resources.

\paragraph{Threats to validity.} 
We first would like to emphasize that our goal is not a systematic evaluation of all the 30 papers of the prevalence study across all the hazards, but more importantly to assess the impact of the hazards on the claims of the papers. Thus, we need at least for each hazard one method/case where the taking into account the hazard yields different conclusions. To cover our ten hazards, the six testing methods we evaluated were sufficient.

We restricted our empirical study only to the approaches with available public implementation. We relied on the library DNN-TIP \cite{dnntipp} for the Pytorch implementation of Neuron Coverage and Surprise Adequacy, and adapted the gradients steps of the remaining approaches from TensorFlow to PyTorch. Adversarial training of our large models is expensive, hence we used mixed precision training to speed up the process. These changes may cause slight differences compared to the original approaches. Besides, while our search and paper collection was very careful, we may have missed work in preprint at the time. Finally, our list of hazards comes from common pitfalls we identified in the collected papers. Additional hazards may be relevant to specific settings. 

We focused on image classification testing for two main reasons: First, across the papers on the study, only three (DeepXplore \cite{pei2017deepxplore}, SENSEI\cite{sensei}, and CoEva2\cite{ghamizi2020search}) extended their evaluation outside of image datasets, second well established benchmarks of robustness and generalization are only available for image classification with RobustBench\cite{croce2020robustbench}. We could have picked robust models in NLP from specific approaches, but their protocol quality and their validity would be lower than relying on pretrained robust models well recognized and evaluated by the community.  

\section*{Conclusion}

We have highlighted 10 critical hazards that may negatively impact the validity of ML testing experiments and yield overestimated performance of ML models and systems. These pitfalls are prevalent in this research area and affect even the most popular methods. To assist researchers in avoiding these pitfalls, we have provided recommendations that can be applied to generate new testing inputs or to assess the quality these inputs. 

Our aim is to improve the scientific quality of empirical work on machine learning from a software engineering perspective. We hope this empirical study will increase the awareness of researchers in the recommended experimental practices for ML testing and model evaluation in general.

We provide an anonymized \textbf{replication package} on FigShare\footnote{\href{https://figshare.com/projects/On_Testing_Deep_Learning_Models_Pitfalls_and_Good_Practices_-_Replication_package/163429}{\url{https://figshare.com/projects/On\_Testing\_Deep\_Learning\_Models\_Pitfalls\_and\_Good\_Practices\_-\_Replication\_package/163429}}} that contains the listing of the evaluated papers, the filtering steps and the source code of the empirical study.

\newpage
\clearpage

\begin{acks}

\end{acks}

\bibliographystyle{ACM-Reference-Format}
\bibliography{bib/refs, bib/adv, bib/general, bib/testing}


\end{document}